\documentclass[12pt,preprint]{aastex}

\newcommand{\dxdy}[2]{{\frac{\partial{#1}}{\partial{#2}}}}

\shorttitle{Anelastic Penetrative Convection}
\shortauthors{Rogers et al.}

\begin{document}

\title{Numerical Simulations of Penetration and Overshoot in the Sun}

\author{Tamara M. Rogers} 
\affil{Astronomy and Astrophysics Department, University of California,
    Santa Cruz, CA 95064}
\email{trogers@pmc.ucsc.edu}
\author{Gary A. Glatzmaier}
\affil{Earth Sciences Department, University of California, Santa Cruz, 
	CA 95064}
\author{C.A. Jones}
\affil{Department of Applied Math, University of Leeds, Leeds LS2 9JT, UK}

\begin{abstract}
We present numerical simulations of convective overshoot in a two-dimensional model of the solar equatorial plane.  The simulated domain extends from 0.001 R$_{\odot}$ to 0.93 R$_{\odot}$, spanning both convective and radiative regions.  We show that convective penetration leads to a slightly extended, mildly subadiabatic temperature gradient beneath the convection zone below which there is a rapid transition to a strongly subadiabatic region.  A slightly higher temperature is maintained in the overshoot region by adiabatic heating from overshooting plumes.  This enhanced temperature may partially account for the sound speed discrepancy between the standard solar model and helioseismology.  Simulations conducted with tracer particles suggest that a fully mixed region exists down to at least 0.687 R$_{\odot}$.  

\end{abstract}


\section{Introduction}

One of the main unsolved problems in all of stellar evolution theory is the treatment of convective-radiative boundaries.  In the Sun, understanding convective overshoot is crucial in determining properties of the solar tachocline and the solar dynamo.  Helioseismic observations have shown that the differential rotation observed at the solar surface persists throughout the convection zone (Thompson et al. 1996).  More surprisingly, the p-mode splittings showed that over a very thin radial extent the rotation profile changes from differential in the convection zone, to solid body in the radiative interior.  The unresolved radius over which this transition occurs has been named the tachocline (Spiegel \& Zahn 1992).  Whether this region is quiescent and mostly devoid of overshooting motions or is violent and constantly bombarded by plumes is an unanswered question, the solution to which may help constrain theoretical models for the tachocline.  

Understanding the tachocline is not only important for comprehending the internal rotation of the Sun; it is crucial for understanding the dynamo process.  In classic mean field theory, the dynamo process is explained in two steps: poloidal field is sheared into toroidal field by differential rotation (the $\Omega$ effect), toroidal field then buoyantly rises and, because of Coriolis forces is twisted back into a poloidal field (the $\alpha$ effect).  These processes were initially postulated to occur in the convection zone.  However, it became clear some time ago (Parker 1975) that magnetic field would become buoyant and be quickly shredded within the turbulent convection zone.  During the ensuing years it was proposed that the storage and amplification of field could occur in the stably-stratified overshoot region (Spiegel \& Weiss 1980).  The tachocline, with its strong differential rotation would provide the ideal site for storage and amplification of the field.  To explain how the $\alpha\Omega$ dynamo would work in this scenario Parker (1993) proposed the interface dynamo, which places the $\Omega$ effect in the stable region beneath the convection zone, while keeping the alpha effect in the bulk of the convection zone.  If the interface dynamo is to work, the overshoot region must play two crucial roles in the dynamo cycle: (1) the stable stratification allows field, which is transported into the region by overshooting plumes (Tobias et al. 1998), to be stored there on the solar cycle timescale and (2) the strong shear in this region must convert poloidal field into toroidal field.  Understanding the precise nature of this overshoot region, the amplitude of the subadiabaticity and its depth, is crucial for understanding the efficiency with which field can be pumped into the overshoot region, the magnitude of the field capable of being stored there and the timescale on which it could be stored.   

The problem of overshoot is not confined to the Sun.  Most stages of stellar evolution are affected.  The Lithium depletion in some main sequence stars may be explained by convective overshoot.  In more massive stars with convective cores, overshoot can lead to increased central hydrogen abundance, and therefore longer main sequence lifetimes, which in turn affects isochrone fitting and age predictions for stellar clusters.  Dredge up during the Asymptotic Giant Branch (AGB) phase can lead to the surface enrichment of material from the core and to the presence of $^{13}C$ necessary for s-process nucleosynthesis (Herwig 2000).  Convective overshoot can affect the energy production and nucleosynthesis of classical novae by enriching the accreted layer with underlying white dwarf material (Woosley 1986).

Clearly, overshoot is an important physical process which must be explained if stellar evolution is to be better understood.  Many analytic and numerical theories have been proposed, but no clear consensus has been reached.  Early analytic models (Shaviv \& Salpeter 1973, van Ballegoojen 1982, Schmitt, Rosner \& Bohn 1984) predicted extended adiabatic regions and early numerical experiments concurred (Hurlburt et al. 1994).  More recent analytic results (Zahn 1991, Rieutord \& Zahn 1995, Rempel 2004) model the penetration using plume dynamics and agree that the extent and nature of the overshoot depends critically on the filling factor of the plumes at the base of the convection zone as well as the flux.  It also appears that the inclusion of upflow-downflow interaction, and the prescription for it, affects the penetration depth (Rieutord \& Zahn 1995, Rempel 2004).  More recent numerical experiments at higher spatial resolution in two- (Rogers \& Glatzmaier 2005a) and three dimensions (Brummell et al. 2002) have found subadiabatic overshoot but no extended adiabatic region.   

All analytic predictions make some crude assumptions, while all numerical simulations employ far from realistic solar parameters.  The numerical simulations presented in this work still can not reach the level of turbulence that surely exists in the Sun.  However, in this work we have taken a few steps forward by using a realistic thermal profile, including rotation, employing a more realistic geometry and including most of the convective and radiative regions.  In section 2 we describe our numerical model and equations; in section 3 we discuss some of the features of basic convective penetration.  In section 4 we present estimates for the depth of the penetrative convection before our reference state model is evolved.  Section 5 details the evolution of the mean thermal profile and section 6 compares these results with previous analytic and numerical models.  In section 7 we discuss mixing of species as modeled by tracer particles.

\section{Numerical Model}
\subsection{Equations}
The numerical technique used here is identical to that in Rogers \& Glatzmaier (2005b).  We solve the Navier Stokes equations in the anelastic approximation (Gough 1969) for perturbations about a mean thermodynamic reference state.  The equations are solved in 2D cylindrical geometry, with the computational domain extending from 0.001 R$_{\odot}$ to 0.93 R$_{\odot}$, representing the equatorial plane of the Sun.  The radially (r) dependent reference state is taken from a standard 1D solar model (Christensen-Dalsgaard private communication).  The radiative region spans from 0.001R$_{\odot}$ to 0.718R$_{\odot}$, while the convection region occupies the region from 0.718R$_{\odot}$ to 0.90R$_{\odot}$ and for numerical ease an additional stable region is included from 0.90R$_{\odot}$ to 0.93R$_{\odot}$.  Excluding 0.1 \% of the solar radius in the core should have little effect, as gravity waves reflect when their frequency approaches the Brunt-Vaisala frequency, which vanishes at the core.\footnote{However, we do not fully understand the properties of nonlinear waves under internal reflection.  It is possible that internal reflection is physically distinct from reflection off of a hard boundary.}

For numerical simplicity we solve the curl of the momentum equation, the vorticity equation:

\begin{equation}
\dxdy{\omega}{t}+(\vec{v}\cdot\vec{\nabla})\omega=(2\Omega + \omega)h_{\rho}v_{r}-\frac{\overline{g}}{\overline{T}r}\dxdy{T}{\theta}-\frac{1}{\overline{\rho}\overline{T}r}\dxdy{\overline{T}}{r}\dxdy{{\it{p}}}{\theta} +\overline{\nu}\nabla^{2}\omega
\end{equation}
where, $\omega=\vec{\nabla}\times\vec{v}$ is the vorticity in the $\hat{z}$ direction and $\vec{v}$ is the velocity, comprised of a radial component, $v_{r}$, and a longitudinal component,  $v_{\theta}$.  T is the temperature, $\rho$ is the density, ${\it p}$ is the pressure, and g represents gravity.  Overbars denote prescribed reference state variables taken from a 21st order polynomial fit to the 1D solar model.  These values are functions of radius and slow functions of time when the reference state is allowed to evolve (see section 2.2).  Variables without overbars are the time dependent perturbations which are solved for relative to the reference state.  Derivatives of the reference state values are calculated from analytic derivatives of the polynomial expansions.  The viscous diffusivity, $\overline{\nu}$, is radially dependent and defined such that the dynamic viscosity, $\overline{\nu\rho}$ is constant (initially).  The rotation rate, $\Omega$, is set equal to the mean solar rotation rate, 2.6$\times 10^{-6}$ rad/s.  

The energy equation is solved as a temperature equation: 
\begin{eqnarray}
\lefteqn{\dxdy{T}{t}+(\vec{v}\cdot\nabla){T}=-v_{r}(\dxdy{\overline{T}}{r}-(\gamma-1)\overline{T}h_{\rho})+}\nonumber\\
&  & {(\gamma-1)Th_{\rho}v_{r}+\gamma\overline{\kappa}[\nabla^{2}T+(h_{\rho}+h_{\kappa})\dxdy{T}{r}]+}\nonumber\\
&  & \gamma\overline{\kappa}[\nabla^{2}\overline{T}+(h_{\rho}+h_{\kappa})\dxdy{\overline{T}}{r}] + \frac{\overline{Q}}{c_{v}}
\end{eqnarray}
In Equation (2), $\overline{\kappa}$ is the (radially dependent) thermal diffusivity, $(\gamma-1)=(dln\overline{T}/dln\overline{\rho})_{ad}$, $h_{\rho}=dln\overline{\rho}/dr$ and $h_{\kappa} = dln\overline{\kappa}/dr$.  These last two, h$_{\rho}$ and h$_{\kappa}$, represent inverse scale heights.  $\overline{Q}$ represents physics included in the 1D standard solar model, but not included here, which maintains the initial reference state temperature gradient.  In the convection zone, $\overline{Q}$ is the divergence of the mixing length flux which, together with the second to last term in (2) accounts for the convergence of the total reference state flux through the system.  Initially, this sum is set to zero, so that the initial reference state temperature is time independent.  

In this model, similar to Rogers \& Glatzmaier 2005, we use the temperature as our working thermodynamic variable, rather than entropy.  We do this to avoid the inward heat flux that would accompany a positive entropy gradient and large turbulent diffusivity in the stable, radiative interior.  
 
We calculate the pressure term in (1) by solving the longitudinal component of the momentum equation: 
\begin{equation}
\frac{1}{\overline{\rho}r}\dxdy{\it{p}}{\theta}=-\dxdy{v_{\theta}}{t}-(\vec{v}\cdot\vec{\nabla}\vec{v})_{\theta}+\overline{\nu}[(\nabla^{2}\vec{v})_{\theta} - \frac{h_\rho}{3r}\dxdy{v_{r}}{\theta}] 
\end{equation}
These equations are supplemented by the continuity equation in the anelastic approximation: 
\begin{equation}
\nabla \cdot \overline{\rho} \vec{v} = 0 
\end{equation}

The convective and radiative regions are set up by taking the subadiabaticity defined as:
\begin{equation}
\Delta\nabla T =(\overline{\gamma} -1)h_{\rho}\overline{T}-\dxdy{\overline{T}}{r}\approx ((\dxdy{\overline{T}}{r})_{ad} -\dxdy{\overline{T}}{r})
\end{equation}
from the 1D model and specifying the superadiabaticity in the convection zone as the constant value $10^{-7}$ K/cm.  The thermal diffusivity, $\overline{\kappa}$ is given by the solar model, multiplied by a constant factor, $\kappa_{m}$ for numerical stability:
 
\begin{equation}
\overline{\kappa}=\kappa_{m}\frac{16\sigma \overline{T}^{3}}{3\overline{\rho}^{2}\overline{k}c_{p}}
\end{equation}
where $\sigma$ is the Stefan-Boltzman constant, $\overline{k}$ is the opacity (taken from the 1D solar model) and c$_{p}$ is the specific heat at constant pressure.  The multiplying factor, ${\kappa_{m}}$, is $10^{5}$ and therefore, the convective heat flux is $10^{5}$ larger than the solar value.  This produces the proper radial profile of the radiative diffusivity, albeit increased by a large factor for numerical stability; this is a ``turbulent'' diffusivity. 

Since both $\overline{\nu}$ and $\overline{\kappa}$ vary with height, the relevant control parameters, such as Pr (Prandtl number = $\overline{\nu}/\overline{\kappa}$), Ra (Rayleigh number =$\overline{g} \Delta\nabla T D^{4}/\overline{\nu}\overline{\kappa}\overline{T}$, where D represents the convection zone depth) and Ek (Ekman number =$\overline{\nu}/(2\Omega D^{2})$) vary with height.  The Pr varies from $10^{-2}_{BCZ}$ (at base of the convection zone) to $0.7_{TCZ}$ (at top of convection zone) and is $10^{-3}$ near the core.  The Ra varies from $\approx 10^{8}_{BCZ}$to $\approx 10^{7}_{TCZ}$ and the Ekman number (Ek=$\bar \nu /2 \Omega D^{2}$) varies from $10^{-4}_{BCZ}$ to $10^{-2}_{TCZ}$. 

These equations are solved using a Fourier spectral transform in the longitudinal ($\theta$) direction and a finite difference scheme on a non-uniform grid in the radial (r) direction.  Time advancing is done using the explicit Adams-Bashforth method for the nonlinear terms and an implicit Crank-Nicolson scheme for the linear terms.  The domain resolution is 2048 x 1500, with 620 radial levels dedicated to the radiative region.  The radial resolution in the overshoot region is 170km.  The boundary conditions on the velocity are stress-free and impermeable, while the temperature boundary condition is specified as constant temperature gradient at the top and constant temperature at the bottom.

\subsection{Updating the Reference State}

After the model had run $\approx$ 1 year (which requires nearly 4 million timesteps), we allow the mean reference state to evolve in response to the convection.  The procedure is as follows:
 
(1) Half of the mean temperature, density and pressure perturbations (${\it T(m=0,r)}$, ${\it \rho(m=0,r)}$ and ${\it p(m=0,r)}$) are added to the reference state values $\overline{T}$, $\overline{\rho}$, $\overline{p}$ forming the new reference state:
\begin{equation}
\overline{T}_{new}=\overline{T}_{old}+\frac{1}{2}{\it T(m=0,r)}
\end{equation}
\begin{equation}
\overline{\rho}_{new}=\overline{\rho}_{old}+\frac{1}{2}{\it \rho(m=0,r)}
\end{equation}
\begin{equation}
\overline{P}_{new}=\overline{P}_{old}+\frac{1}{2}{\it P(m=0,r)}
\end{equation}
Here, m is the longitudinal wavenumber and m=0 represents the axisymmetric (mean) perturbation.  The remainder of the mean perturbation is maintained as a perturbation so that the model can adjust gradually.

(2) Using the new values for density and temperature, a new opacity is calculated using Kramers Law
\begin{equation}
k(r)=C(r) \overline{\rho}_{new}(r) \overline{T}_{new}(r)^{-3.5}
\end{equation}
where C(r) is obtained by matching the original opacity profile, taken from the 1D solar model, to a Kramer's Law opacity, i.e. C(r)$\rho_{old}(r)T_{old}(r)^{-3.5}=k(r)_{1Dsolarmodel}$.  

(3) The thermal diffusivity $\overline{\kappa}$ is recalculated with the new opacity, temperature and density via equation (6), the multiplying factor ${\kappa_{m}}$ is maintained.

(4) The last two terms in (2) are updated, holding $\overline{Q}$ constant.

Initially, this procedure is done at regular intervals.  After the initial relatively large changes, this procedure is done only if the mean thermodynamic perturbations become larger than 1\% of the reference state values.

\section{Convective penetration and Gravity Wave generation}

Turbulent convection is dominated by plumes, which form near the top of the convection zone and descend through the bulk of the convection zone.  These plumes tilt and sway and often merge with neighboring plumes during their descent.  Within the convection zone the kinetic energy spectra varies between $m^{-2}$ and $m^{-4}$ depending on the radius at which it is measured and the Ekman number.  The plumes terminate their descent upon encountering the stiff underlying stable region (Figure 1).  In this region a downwelling plume adiabatically heats relative to the subadiabatic background (hence the white ``spots'' seen in Figure 1 at the base of the convection zone) and is then rapidly decelerated by buoyancy.  The depth over which the plume is buoyantly stopped depends sensitively on the stiffness of the underlying radiative region (Hurlburt et al. 1994, Brummell et al. 2002, Rogers \& Glatzmaier 2005a).  

When the plumes impinge on the underlying radiative region they generate a spectrum of gravity waves, ranging in frequency from 1$\mu$Hz to 300$\mu$Hz (Rogers \& Glatzmaier 2005b).  Higher frequency waves set up standing modes which may ultimately be detected by helioseismology.  The lower frequency waves are radiatively dissipated, thereby sharing their angular momentum with the mean flow.  

\section{Convective Overshoot}

There are several ways to define the depth of convective overshoot.  In this section we consider only the depth to which subadiabatic overshooting motions can extend into the stable radiative region.  In the next section we consider whether these motions can extend the adiabatic region beyond that prescribed by the reference state model.  As we are concerned with the cessation of large amplitude motion, we consider the amplitudes of kinetic energy density $\rho(v_{r}^{2}+v_{\theta}^{2})$, vertical kinetic energy flux $v_{r}$(kinetic energy density) and convective heat flux ($\overline{\rho} Tv_{r}c_{p}$) as a means for determining the depth of convective overshoot.  The time and horizontally averaged kinetic energy density as a function of radius is shown in Figure 2a.  The kinetic energy density drops rapidly moving into the radiative region because of the stiff underlying stable region.  The peak energy density is $\approx 5\times 10^{8}$ergs/cm$^3$, but drops to $10^{4}$ ergs/cm$^{3}$ at just one H$_{p}$ (pressure scale height) below the convection zone \footnote{Note that the kinetic energy density presented here is larger than one might expect in the Sun.  This is due to increased velocities driven by a larger than solar thermal diffusivity.}.  If we define the depth of overshoot to be the distance between the base of the convection zone and the height at which the kinetic energy density reaches 1\% of its peak value, then that depth is 5.3$\times10^8$cm, or 0.09 H$_{p}$ (H$_{p}$ is the pressure scale height at the base of the convection zone, $5.8\times10^{9}$cm).  If instead we use 5\% of the peak, the depth becomes 0.06 H$_{p}$.  

An alternative, more physical way of defining the depth of convective overshoot is based on where the kinetic energy flux changes sign.  The average vertical kinetic energy flux\footnote{Plotted in Figure 2b is the kinetic energy flux divided by the reference state diffusive heat flux, $\overline{\kappa\rho} c_{p} d\overline{T}/dr$.} (Figure 2b) is negative (downward) in the bulk of the convection zone due to descending plumes.  In this model, because of the stiffness of the underlying radiative region, the flux changes sign well within the convective region and is positive at the base of the convection zone.  In this region descending plumes rebound, causing a net upward flux.  The extent of the region with positive vertical kinetic energy flux could be considered an overshoot depth.  This yields a smaller depth of  0.02 H$_{p}$.  

Another physical measure of overshoot can be found if one considers the correlation between temperature perturbation and vertical velocity i.e., the convective heat flux.  Typically, a positive temperature perturbation (hot fluid) would give rise to a positive vertical velocity.  However, in the overshoot region, positive temperature perturbations are associated with negative vertical velocities.  Therefore, in the overshoot region the quantity $Tv_{r}$ is negative, as seen in Figure 2c, leading to a similar measure of the overshoot depth, 0.03H$_{p}$.  The reduction in kinetic energy density from peak is 20\% for 0.03H$_{p}$ penetration depth and 28\% for 0.02H$_{p}$.  

\section{Evolving the Reference State}

In order to study how the overshooting motions discussed above affect the mean thermal state of the system, we allow the model to evolve in response to those motions.  We can then determine if we find an extended adiabatic region or just subadiabatic overshoot.  An extended adiabatic region is generally referred to as ``penetration depth'' and distinguished physically and colloquially from overshoot, or just subadiabatic overshoot.  After the model has run nearly one year the reference state thermodynamic variables are updated.  The horizontal average of the temperature perturbation (i.e. the m=0 mode of the Fourier expansion) is added to the reference state temperature, as described in section 2.2.  Similarly, the horizontally averaged density and pressure perturbations are added to their reference state values.  The newly formed density and temperature are used to define a new opacity using Kramers Law (10).  Using the new opacity and thermodynamic variables a new thermal diffusivity is obtained using (6).  A new reference state is thus formed.  Initially, this procedure is done periodically.  Later, the reference state is changed only when the temperature perturbation reaches 1\% of the mean temperature at any location.  

At any given instant the temperature perturbations are a small fraction of the mean temperature; however, the continual evolution of the temperature, density and opacity produce large changes in the mean thermal profile (Figure 3).  In Figure 3a we show the super-/subadiabaticity, as defined in (5), for the initial model (solid line) and our evolved model (dotted line); in Figure 3b we show the temperature.  We show only a small area around the overshoot region so as to highlight the regions where the most change has occurred\footnote{There is very little change in the deep radiative interior, nor in the bulk of the convection zone.}.  The bulk of the convection zone remains superadiabatic, however, near the base of the convection zone, the region becomes slightly subadiabatic.  In addition, just below the convection zone there is a slightly extended mildly subadiabatic region, where the initial steep gradient between super- and subadiabatic temperature gradients has evolved into a smoother transition.  In this region the temperature gradient (see Figure 3b, region labeled 1) becomes slightly steeper so that the upward diffusive heat flux can increase to compensate for the negative convective heat flux in that region (Figure 2b).  The depth of the extended mildly subadiabatic region is roughly 0.05 H$_{p}$, if that depth is measured at the point where the subadiabaticity reaches $1\times10^{-5}$ (which is marked by the arrow in figure 3a).  However, the choice of this depth is arbitrary as it is unclear where the model changes from ``mildly subadiabatic'' to ``strongly subadiabatic''.

Convective motions are continually pumping heat into the region between labels 1 and 2 in Figure 3b.  The timescale for this transfer is much shorter than the diffusive timescale and therefore, heats up relative to the standard solar model.   This mild heating causes a flatter temperature profile at the region labeled 2 and a steeper profile at region 1.  Hence, region 1 becomes less subadiabatic and region 2 becomes more subadiabatic, which makes the change from nearly adiabatic to strongly subadiabatic occur over a shallower depth than in the standard solar model.  

This small difference in temperature between the standard solar model and our evolved, hydrodynamic model is in the same sense, and at the same radius, as the standard solar model-helioseismology sound speed discrepancy (Christensen-Dalsgaard 2002).  In region 2 the maximum difference in temperature as a fraction of the original temperature ($\delta T/T$) is $\sim$0.019.  This difference has approximately the same amplitude and is in the same direction as the previous standard solar model-helioseismology discrepancy, before gravitational settling was included, but is larger than that discrepancy when gravitational settling is included.  It is possible this effect provides an additional explanation for the sound speed anomaly between the standard solar model and helioseismology.

Because of the extended, mildly subadiabatic region seen in Figure 3, convective motions can penetrate further into the stable region (Figure 4).  In Figure 4 we see that substantial kinetic energy density, kinetic energy flux and convective heat flux stretch further into the stable region compared with Figure 2 (before updating the background state).  Within the convection zone, the now non-constant superadiabaticity yields kinetic energy flux and convective heat flux which are not as smooth as they were previously.  Key physical features, such as a positive kinetic energy flux and negative convective heat flux, just below the convection zone, are still apparent.  Subadiabatic overshooting motions in an evolved background state can extend down to 0.38 H$_{p}$ (0.687 R$_{\odot}$), still significantly above the base of the tachocline at $\sim$ 0.65 R$_{\odot}$.  In summary, we find an extended ${\it mildly}$ subadiabatic region down to a depth of roughly 0.05H$_{p}$\footnote{Measured at the arrow in figure 3.6a with the stipulation that the definition of ${\it mildly}$ subadiabatic is subjective.}, with subadiabatic overshoot extending further to 0.38H$_{p}$.
    
\section{Comparison with Previous Results}

A review of literature on the subject of penetrative convection yields various predictions and no consensus.  Both numerical and analytic models have been used.  Early numerical simulations in two dimensions (Hurlburt et al. 1994) found that the depth of penetration depended primarily on the ratio of the sub- to superadiabaticity, S.  These simulations found that for low values of this ratio (1-4), the penetration depth scaled as $S^{-1}$ indicating an extended adiabatic region.  For larger values of S (between 5 and 20), the penetration scaled as d$_{pen} \propto S^{-1/4}$.  Simulations of more turbulent convection in 3D (Brummell et al. 2002) found only strongly subadiabatic overshoot and no extended adiabatic region, hence retrieving only the d$_{pen} \propto S^{-1/4}$ scaling.  The authors argue this discrepancy could be due to the much smaller filling factor that arises in the 3D simulations or the large velocities attained in 2D due to flywheel type motion.  Our more recent simulations in 2D, using stiffer values for the stability (closer to solar values) of the underlying radiative regions find no extended adiabatic region (Rogers \& Glatzmaier 2005).  The discrepancies between the different numerical simulations can be traced to the reference state employed (in particular, the value of subadiabaticity), the degree of turbulence and the dimensionality of the model.  

Analytic models have been in somewhat better agreement.  Early models by Shaviv \& Salpeter (1973), van Ballegoojen (1982) and Schmitt, Rosner \& Bohn (1984) all found penetration depths ranging from 20\% to 40\% H$_{p}$ \footnote{Note that this is the extent of the extended mildly subadiabatic region, not the extent which plumes overshoot (subadiabatic overshoot).}.  This appeared to be a robust solution as they all arrived at these values using vastly different models.  The work by Schmitt, Rosner \& Bohn (1984) was the first of the modern analytic solutions which model penetration using plumes (Zahn 1991, Rieutord \& Zahn 1995, Rempel 2004).  All of these plume models find that the penetration depth depends critically on the filling factor, ${\it f}$, that is, the fractional area occupied by the plumes at the base of the convection zone.  Schmitt, Rosner \& Bohn (1984) found that the penetration depth scaled as ${\it v^{3/2}f^{1/2}}$; later, Zahn (1991) provided a derivation of this scaling.  He showed that:
\begin{equation}
d_{pen}^{2}=\frac{3}{5}H_{p}H_{\chi}{\it f} \frac{\rho {\it V^{3}}}{F_{total}}
\end{equation}
demonstrating how the penetration depth depends not only on the filling factor (${\it f}$), but also on the plume exit velocity (V), total heat flux (F$_{total}$) and thermal diffusivity (through H$_{\chi}$).  

If we compare our model directly to the expression derived by Zahn (1991) for the penetration depth (11), we can isolate any fundamental differences.  For instance, both H$_{p}$ and $\rho$ are taken from the solar model in our simulations and therefore, can not be criticized as unrealistic.  While our flux is increased by $10^{5}$ (${\it \kappa_m}$), our velocities are also larger than one would expect in the solar convection zone, and therefore, the ratio $V^{3}/F_{total}$ is likely not unrealistic.  Finally, the fundamental complaint generally levied against numerical simulations is their enhanced thermal diffusivity required for numerical stability.  At issue is that this enhanced diffusivity allows the overshooting plumes to thermalize with the background thermodynamic state more quickly than they would if the solar thermal diffusivity were used, hence leading to lower overshoot depths.  However, as is seen in (11) the quantity upon which the penetration depth depends is H$_{\chi}$, the radiative conductivity scale height.  While our simulation uses an enhanced thermal diffusivity, the radiative conductivity ${\it scale}$ ${\it height}$, represented by H$_{\chi}$ is taken directly from the solar model and therefore, can not be criticized as unrealistic.  The only factor which may be criticized is the filling factor, which we admit, may be enhanced because we are only modeling two dimensions.

Using a more sophisticated plume model \cite{rz95} found that when the effect of the surrounding upflow is included in their plume model the penetration depth decreases almost linearly with the number of plumes.  When the interaction between upflows and downflows is neglected, the penetration depth is found to be between 0.2 H$_{p}$ and 0.4 H$_{p}$, as found in previous analytic solutions, but which was inconsistent with helioseismic observations \citep{jcd95}.  However, when that interaction is included the penetration depth depends sensitively on the number of plumes.  More plumes results in more upward momentum loading by upwelling fluid.  More recently Rempel (2004) considered a semi-analytic overshoot model based on plumes.  His model also includes the interactions of plumes with upflows and finds the same behavior of decreasing penetration depth with increased upflow-downflow interaction (see his figure 7).  These results illustrate that the penetration depth depends, rather strongly, on the upflow-downflow interaction, which is parametrized in the best analytic models, but self-consistently calculated in numerical simulations.  In light of this it seems probable that at least part of the inconsistency between numerical and analytic models lies in the simplified treatment of upflow-downflow interaction in analytic models.

In addition to the dependence on mixing between upflows and downflows (parametrized as $\alpha$) Rempel finds that the depth of overshoot and  the nature of the transition between convective and radiative zones depends on the ratio of the total flux to filling factor:

\begin{equation}
\phi=F_{total}/(fp_{bc}(p_{bc}/\rho_{bc})^{0.5})
\end{equation}
where f is the filling factor, $p_{bc}$ and $\rho_{bc}$ are the pressure and density at the base of the convection zone, respectively.  He finds that the main difference between numerical and analytic results lies in the values of this ratio; with numerical simulations using values of $\phi\sim 10^{-2}$ (because of their increased fluxes) and analytic models employing values around $10^{-9}$.  The model we present here has $\phi\sim 10^{-3}-10^{-4}$ depending on the filling factor of the model.  Our evolved reference state superadiabaticity (Figure 3a) resembles their model for $\phi\sim 10^{-4}$.  However, it is difficult to make a direct comparison, given the dependence of their model on $\alpha$, although there does appear to be some degree of agreement (slightly subadiabatic base of the convection zone, slightly extended mildly subadiabatic region).\footnote{Note that we present the super/subadiabatic temperature gradient, while they show the dlnT/dlnP, so that the amplitudes are not similar but the profiles of super-/subadiabaticity are similar.}

To recap, combined numerical and analytic models have found that the penetration depth depends mainly on: (1) the subadiabaticity of the underlying stable region, (2) the filling factor at the base of the convection zone and (3) the interaction of upflows and downflows.  Low subadiabaticity leads to larger penetration depths, while high subadiabaticity yields small penetration depths.  Large filling factors lead to large penetration depths and vice versa, depending on the number of plumes which occupy that filling factor.  For a large number of plumes the penetration depth decreases, because of enhanced upflow-downflow interaction, while for a low number of plumes the penetration depth increases.  Our numerical simulations accurately represent the subadiabaticity of the radiation zone and the interaction of upflows and downflows.  However, the filling factor depends sensitively on the properties of the convection (Ra, Re, Pr) and on the geometry and dimensionality of the flow.  Resolving the issue of filling factor remains an important issue, which should be addressed both by analytic and numerical studies.
\section{Tracer Particles}

To study the effect of overshooting motions on the mixing of species, we introduce tracer particles, which are advected by the flow.  These particles are introduced after the reference state has been evolved.  We are particularly interested in estimating how deep below the convection zone the fluid is efficiently mixed.

Five sets of 500 tracer particles were initiated at five different radii, equally distributed in longitude (Figure 5a).  Two sets were initiated in the convection zone (at  0.83 R$_{\odot}$ and 0.79 R$_{\odot}$, represented in Figure 5 by red and blue points, respectively) and 3 sets were initiated in the radiative zone (at 0.705 R$_{\odot}$, 0.690 R$_{\odot}$ and 0.675 R$_{\odot}$, represented in Figure 5 by purple, cyan and black, respectively).  The yellow line represents the base of the convection zone.  Figure 5b shows the distribution of the particles after two convective turnover times.  The particles initiated within the convection zone (red and blue) are distributed throughout the convection zone after two convective turnover times.  Of the particles initiated in the stable region many are pulled into the convection zone.  The sets initiated at 0.705 R$_{\odot}$ and 0.690 R$_{\odot}$ both have particles within the convection zone in steady state.  However, no particles initiated at 0.675 R$_{\odot}$ have made it into the convection zone on this timescale, although some have migrated from their initial positions. 

Figure 6 shows the number of particles, of the two sets initiated in the convection zone, which are pumped into the radiation zone (any radius below 0.718 R$_{\odot}$) over two convective turnover times.  Both sets converge to having approximately 90 particles (or 18\% of the initial 500 particles) within the radiation zone at any given time.  However, most of these particles do not reach any appreciable depth below the convection zone, existing just beneath the transition.  Figure 7 shows the number of particles that started in the convection zone and made it into the radiative region as a function of time (left) and radius (right below the convection zone.  The number of particles drops rapidly below the convection zone, with a nearly linear drop down to 0.705 R$_{\odot}$ and a slightly less rapid drop below that.  The concentrations plotted in Figure 7b are the steady state concentrations; however, the tail of the distribution (the depth at which no particles are found) depends on the initial number of particles, as well as the time simulated, which is very short here.  The minimum radius any particle, initiated within the convection zone, reaches over the two convective turnover times is 0.687 R$_{\odot}$.  While this minimum radius is reached only intermittently, it indicates that over longer timescales a fully mixed region would extend to at least this depth.

Of the three sets of particles initiated within the radiative region, only the particles started at 0.705 R$_{\odot}$ and 0.690 R$_{\odot}$, are pulled into the convection zone.  Figure 8 shows the number of particles as a function of time, for the three sets of particles initiated within the radiative region, which make it into the convection zone (any radius larger than 0.718 R$_{\odot}$).  As can be seen in Figure 9, nearly half of the 500 particles (230, 46\%) initiated at 0.705 R$_{\odot}$ are in the convection zone in steady state.  Of the particles initiated at 0.690 R$_{\odot}$, nearly 40 particles (20 \% of those initiated) are within the convection zone in steady state.  {$\it None$} of the particles initiated at 0.675 R$_{\odot}$ make it into the convection zone during these two turnover times.   The maximum radius that any particle initiated at 0.675 R$_{\odot}$ reaches is 0.705 R$_{\odot}$ during the time simulated.  

The number of particles that are pumped down to the radius 0.690 R$_{\odot}$ (1-2) is significantly lower than the number of particles which are dredged up from this same radius (40) in steady state (compare Figures 7 and 8).  That is, there is an asymmetry between the number of particles dredged up from a particular radius and the number of particles pumped down to that same radius.  This is because downwelling plumes in the convection zone have a small cross sectional area and are therefore unable to entrain many particles during their descent.  However, once plumes reach the radiation zone their surface area is increased as they are decelerated.  These plumes spread transversely and are turned upward.  This ``scooping'' motion, allows a greater number of particles to be dredged up than are pumped down.  In the Sun, where the region is completely mixed on timescales much shorter than the evolutionary timescale, this effect is not important.  However, this asymmetry could be very important in the late stages of stellar evolution of massive stars, when the evolutionary time is closer to the convective turnover time.  This effect could also be crucial in Classical Novae or X-ray bursts.  In these environments the dredge up of underlying heavy material into the accreted layer affects the energetics and nucleosynthesis of the explosion.

\section{Conclusions}

We have presented self-consistent, numerical simulations of convective overshoot in a 2D model of the solar equatorial plane.  This model employs a realistic thermal reference state and evolves in response to convection and overshoot.  We find that the convective penetration leads to a thermal profile which is mildly subadiabatic in a shallow region at the base of the convection zone.  Beneath this mildly subadiabatic region the profile drops precipitously to the extreme subadiabaticity of the radiative interior.  Overshooting plumes cause a mild heating in the overshoot region, leading to a region which has a slightly higher temperature and a greater subadiabaticity than the standard solar model.  The region of increased temperature is at the location and in the sense of the helioseismology-standard solar model sound speed discrepancy and may provide an alternative explanation for that difference.  

Passive tracer particles mix only slightly below the convective-radiative interface over the two turnover times followed.  Over longer times, a fully mixed region can be expected at least down to a radius of 0.687 R$_{\odot}$.  At this radius, the Lithium burning timescale is 8 billion years, a factor of 4 to 5 too long.  However, only two turnover times were simulated with tracer particles and its possible that given longer integration time particles initiated in the convection zone could plunge deeper, thus leading to a shorter burning time.  In order to study this further, longer time integrations are needed.  Furthermore, studies varying the number of particles to determine the minimum radius at which there cease to be particles should be conducted.  

We should bare in mind that these results are 2D and far from the turbulent nature of the Sun.  It is unclear what three dimensional effects would be.  While it seems clear that 3D models have smaller filling factors, it is unclear what this affect will be and it depends not only on the filling factor but the number of plumes accross which that filling factor is distributed.  Furthermore, it is not clear what the effect of meridional circulation or the full Coriolis force would be on the penetration depth and mixing.  Studies on the effect of Rayleigh number (turbulence level) on the penetration depth indicate that increasing the Rayleigh number decreases the penetration depth \citep{bct02, rg05a}.  However, it is unclear how these parametrized results come into play in more realistic models.  The jury is still out and awaits more sophisticated, 3D numerical simulations.

\acknowledgments

We would like to thank P.Garaud, D.O. Gough, J.C. Dalsgaard, Keith MacGregor, N. Brummell and R. Rosner for their guidance and thoughtful insight.  T.R. would like to thank the NPSC for a graduate student fellowship.  Support has also been provided by NASA SHP04-0022-00123 and NASA NAG5-11220 and DOE DE-FC02-01ER41176.  Computing resources were provided by NAS at NASA Ames and by an NSF MRI grant AST-0079757.

\clearpage
\begin{figure}
\plotone{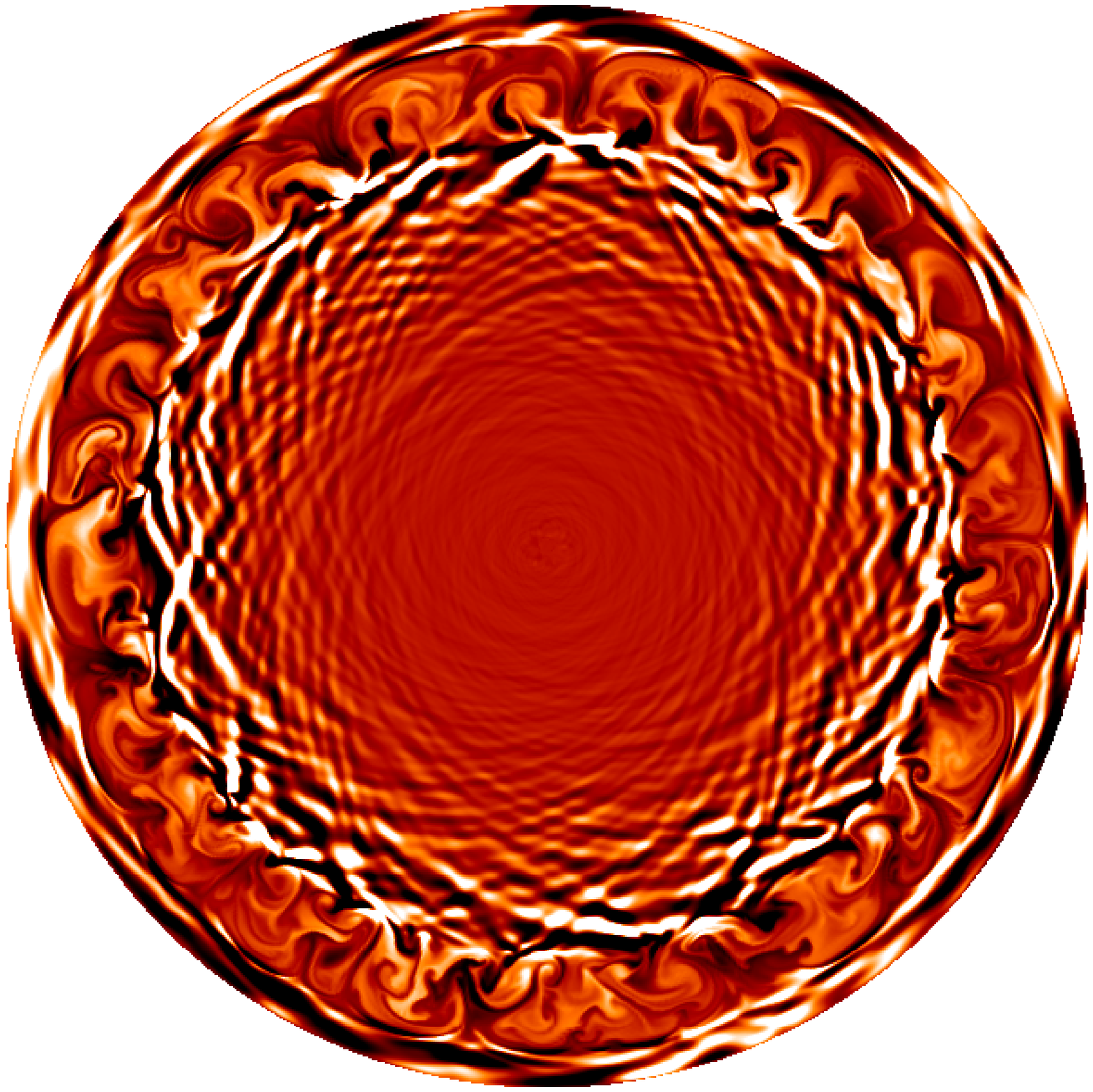}
\caption{Snapshot of the temperature perturbation, representing the full computational domain.  Dark red/white represent cold/hot perturbations with respect to teh background temperature.  The convection region is dominated by descending plumes which overshoot into the radiative region, finding themselves hotter than their surroundings (white spots at base of convection zone).  Gravity waves are generated by these overshooting plumes.}
\end{figure}

\clearpage
\begin{figure}
\includegraphics[width=6in]{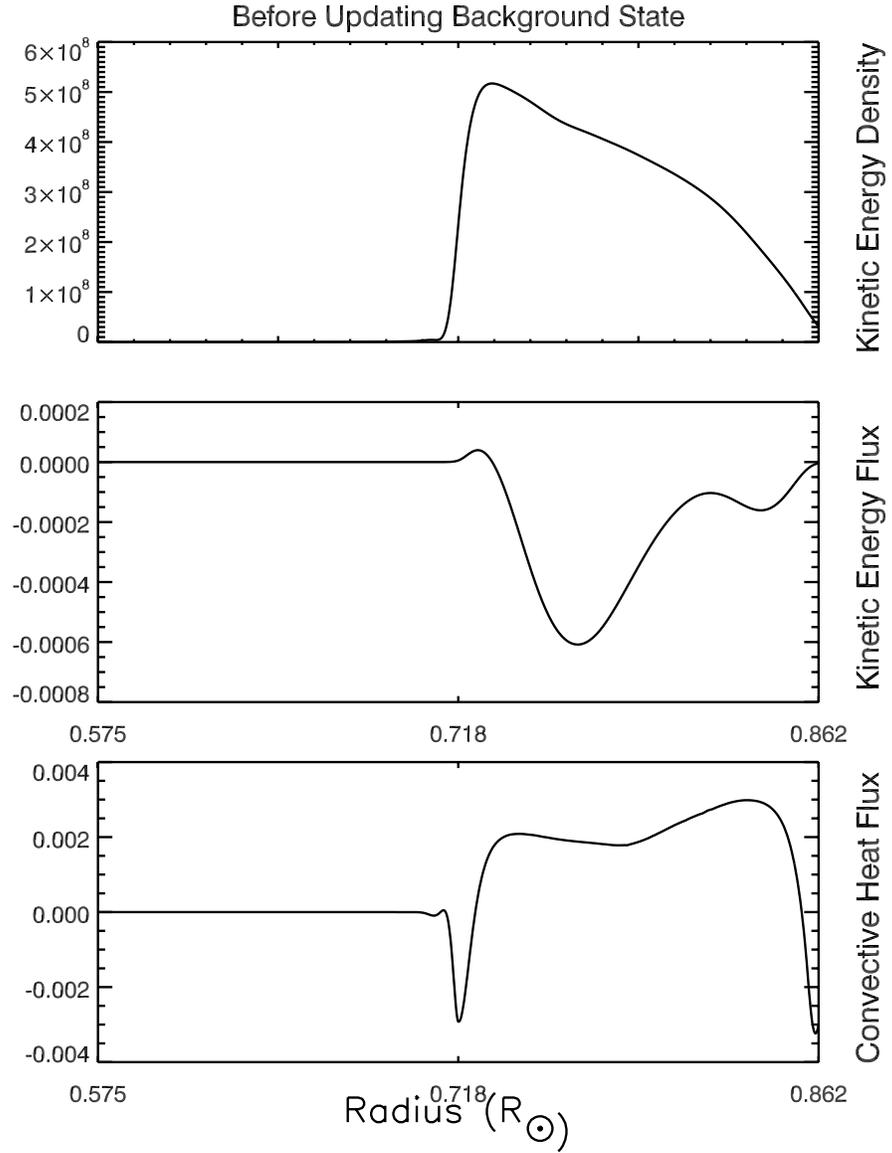}
\caption{Time and longitudinal average of the kinetic energy density (a), vertical kinetic energy flux divided by the reference state diffusive heat flux (b) and convective heat flux divided by the reference state diffusive heat flux(c) before the reference state is updated.  The radius 0.718 R$_{\odot}$ is the interface between the sub- and superadiabatic regions.}
\end{figure}

\clearpage
\begin{figure}
\plotone{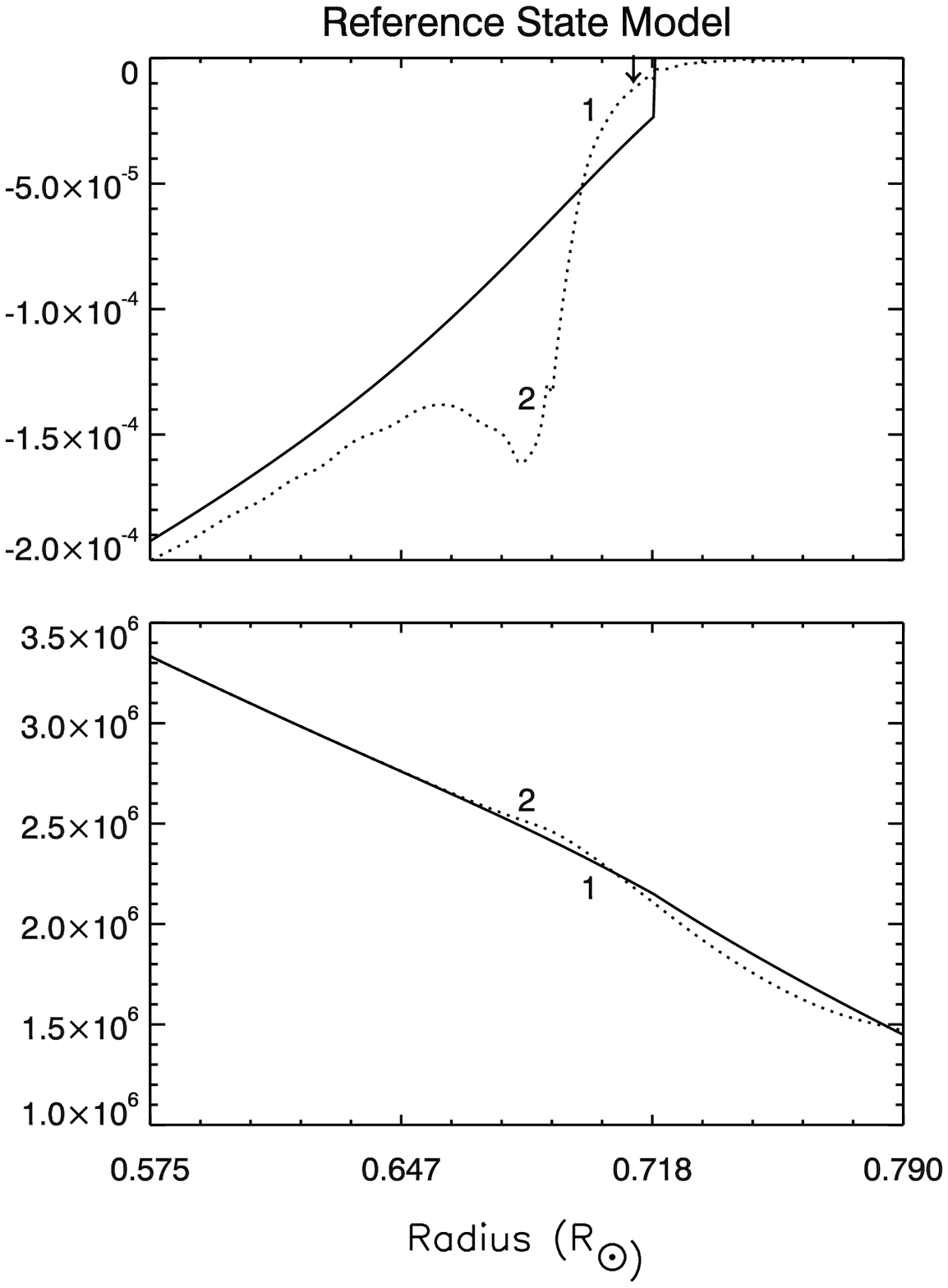}
\caption{Reference state thermodynamic model.  The solid line represents the initial profile; dotted line represents evolved (in response to convection and overshoot) model.  (a) represents the super/sub-adiabatic temperature gradient ) $\Delta\nabla T$ (5) in K/cm, (b) represents the temperature in K. The radius 0.718 marks the transition from convective to radiative zones.}
\end{figure}
\clearpage
\begin{figure}

\includegraphics[width=6in]{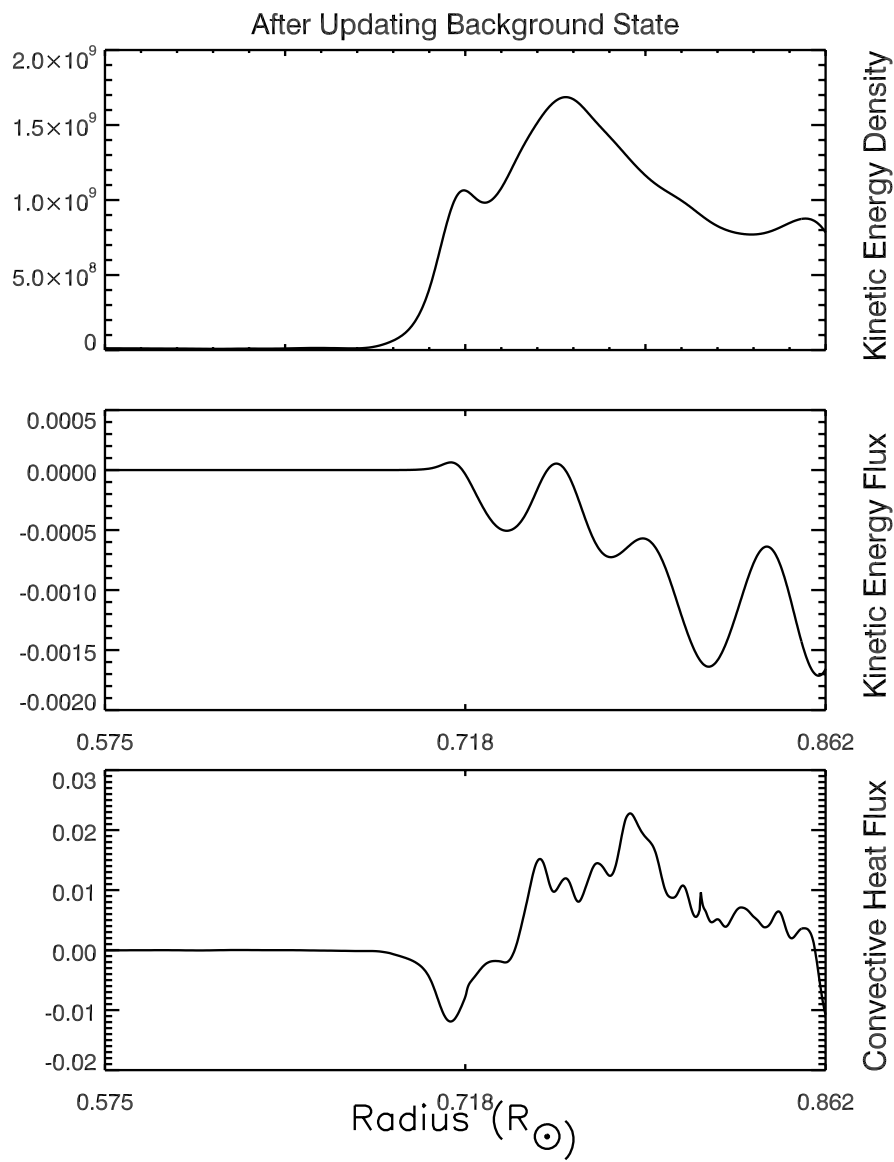}
\caption{Time and longitudinal average of the kinetic energy density (a), vertical kinetic energy flux (b) and convective heat flux (c) after the reference state is updated.  Again, the kinetic energy flux and convective heat flux are divided by the ${\it original}$ reference state diffusive heat flux.  Clearly convective motions penetrate further into the radiative region than in figure 2.  The radius 0.718 was the original interface between the sub- and superadiabatic regions.}
\end{figure}
\clearpage
\begin{figure}
\plotone{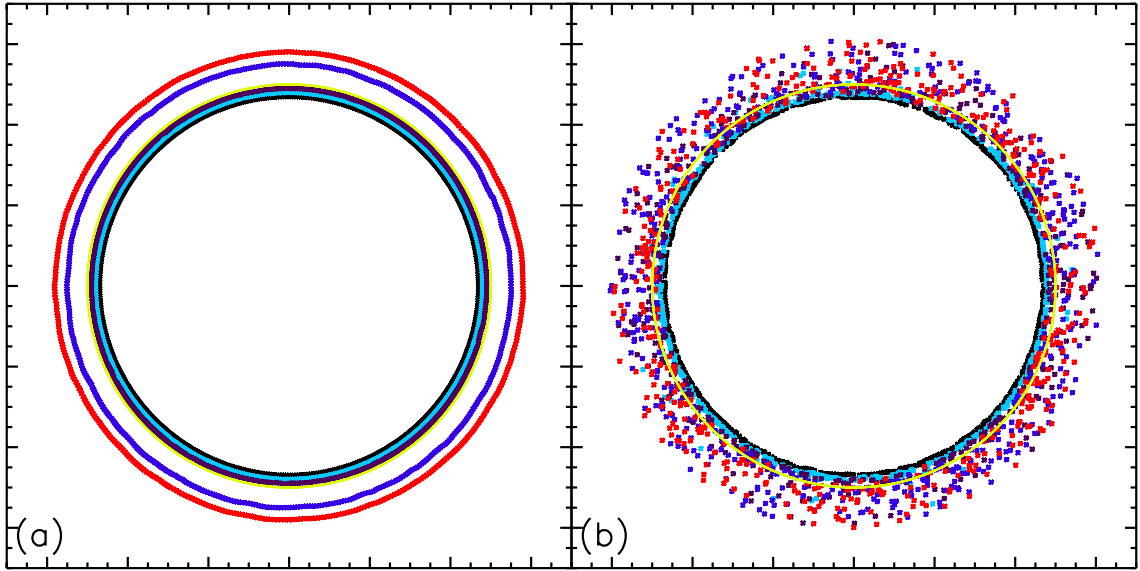}
\caption{Distribution of tracer particles initially (a) and after two convective turnover times (b).  Red and blue particles are initiated within the convection zone, while purple, cyan and black are initiated within the radiative region.  The yellow line represents the convective-radiative interface.}
\end{figure}

\clearpage
\begin{figure}
\includegraphics[width=5in]{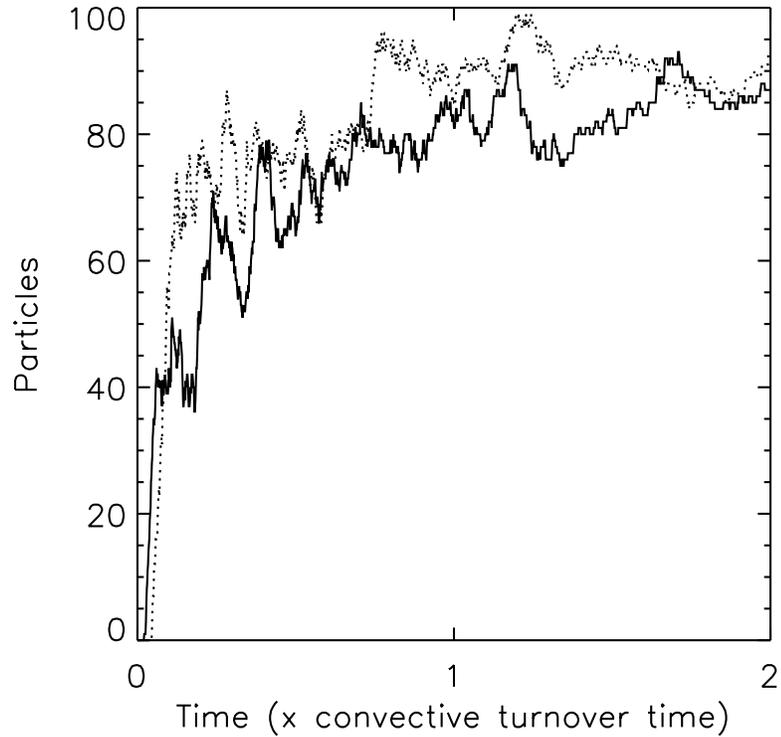}

\caption{Number of particles as a function of time in the radiative region that were initiated within the convection zone.  Solid line represents the particles initiated at 0.833 R$_{\odot}$, while the dotted line represents those particles initiated at 0.790 R$_{\odot}$.  After two convective turnover times the two sets of particles converge to having nearly 90 particles (18\% of those initiated) within the radiation zone in steady state.}
\end{figure}

\clearpage
\begin{figure}
\plotone{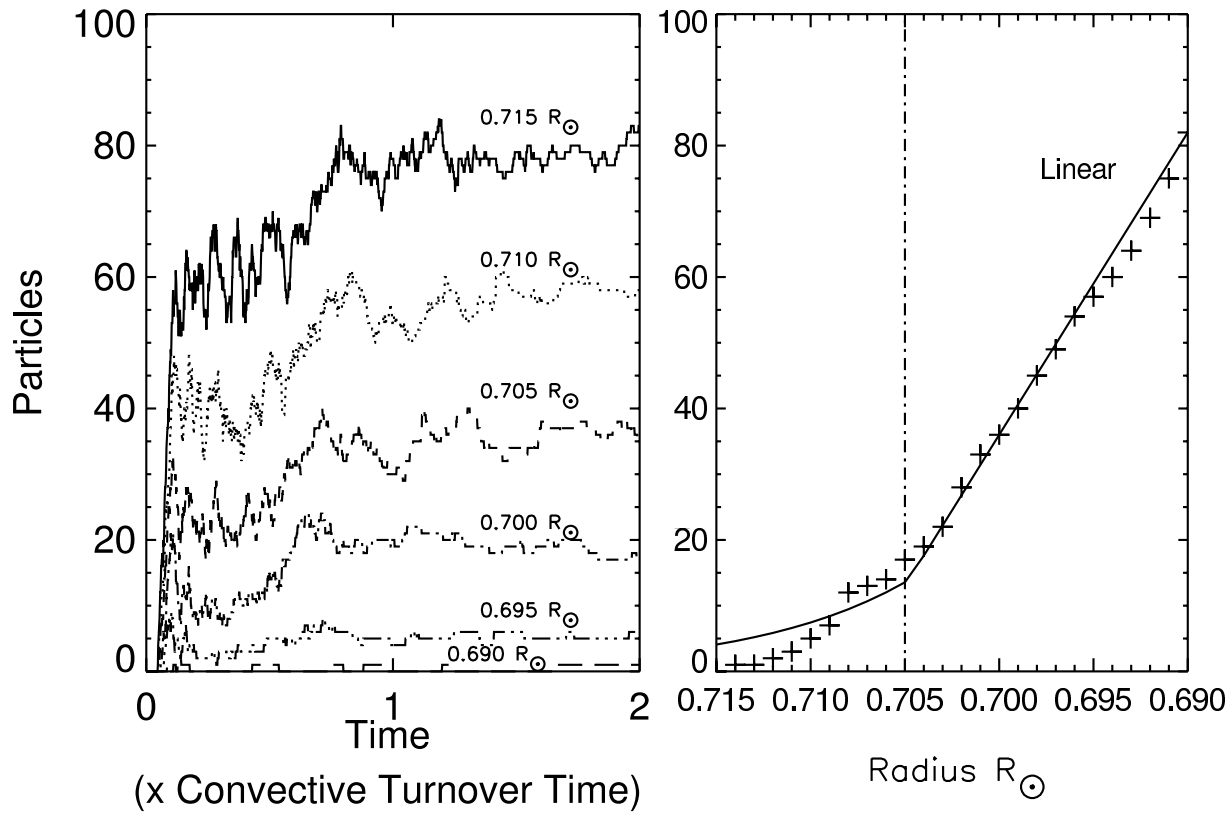}
\caption{(a) Number of particles as a function of time, initiated in the convection zone, that make it to the specified radius within the radiation zone.  (b) Number of particles starting in the convection zone that make it into the radiative region (in steady state) as a function of radius below the convection zone.}
\end{figure}
\clearpage
\begin{figure}
\plotone{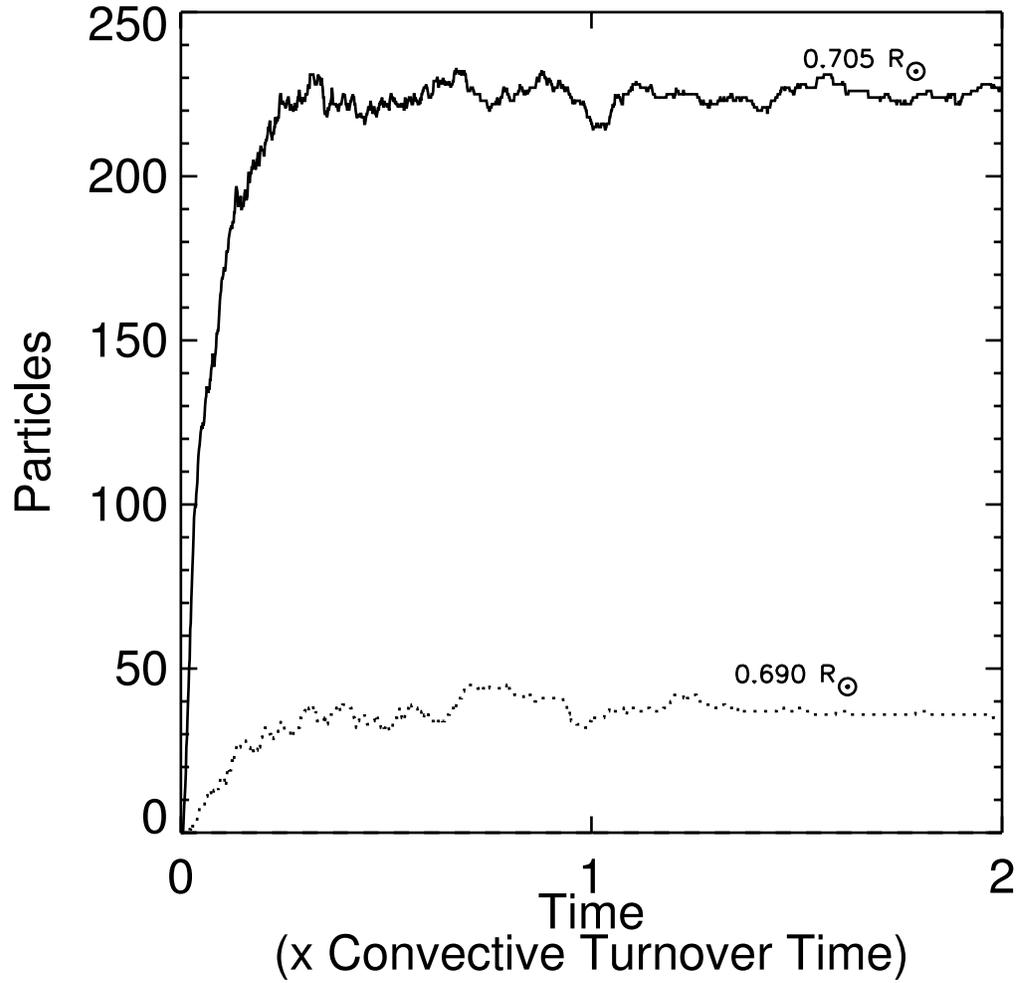}
\caption{Number of particles as a function of time, initiated within the radiation zone (at the specified radii, solid line initiated at 0.705 R$_{\odot}$), dotted line initiated at 0.690 R$_{\odot}$) which make it into the convection zone.  Note, none of the particles initiated at 0.675 R$_{\odot}$ make it into the convection zone during these two convective turnover times.}
\end{figure}








\begin{thebibliography}{}

\bibitem[Brummell et al. (2002)]{bct02} Brummell, N.H., Clune, 
    T.L., Toomre, J. 2002 \apj 570, 825

\bibitem[Cattaneo et al. (1991)]{cat95} Cattaneo, F., Brummell, N., Toomre, J., Malagoli, A. and Hurlburt, N. 1991, \apj, 370, 282

\bibitem[Christensen-Dalsgaard et al (1995)]{jcd95} Christensen-Dalsgaard, J., Mario, J.P.F.G, Monteiro  and Thompson, M.J. 1995, MNRAS 276, 283

\bibitem[Gough (1969)]{go69} Gough, D.O., 1969, J. Atmosp. Sci., 26, 448

\bibitem[Herwig (2000)]{he20} Herwig, F. 2000, \aap, 360, 952

\bibitem[Hurlburt et al. (1986)]{htm86} Hurlburt, N.E., Toomre,
    J. Massaguer, J.M. 1986 \apj, 311, 563

\bibitem[Hurlburt et al. (1994)]{htmz94} Hurlburt, N.E., Toomre,
    J. Massaguer, J.M., Zahn, J.P. 1994 \apj, 421, 245

\bibitem[Massaguer et al. (1984)]{mas84} Massaguer, J. M.,
    Latour, J., Toomre, J., Zahn, J.P. 1984 \aap, 140, 1

\bibitem[Parker (1975)]{pa75} Parker, E.N. 1975, ApJ, 198, 205
\bibitem[Rempel (2004)] {re04} Rempel, M. 2004, \apj, 607, 1046
\bibitem[Rieutord \& Zahn (1995)]{rz95} Rieutord, M. \& Zahn, J.P. 1995, \apj, 296, 127
\bibitem[Rogers \& Glatzmaier (2005a)]{rg05a} Rogers, T.M. \& Glatzmaier, G.A. 2005, \apj, 620, 432
\bibitem[Rogers \& Glatzmaier (2005b)]{rg05b} Rogers, T.M. \& Glatzmaier, G.A. 2005, MNRAS, 364,1135

\bibitem[Schmitt, Rosner \& Bohn (1984)]{srb84} Schmitt, J.H.H.M, Rosner, R., \& Bohn, H.U., 1984, \apj, 282, 316

\bibitem[Shaviv and Salpeter (1973)]{ss73} Shaviv, G. \& Salpeter, E.E. 1973, \apj, 184, 191
\bibitem[Spiegel and Weiss (1980)]{sw80} Spiegel, E.A. \& Weiss, N.O. 1980,\aap, 265,106
\bibitem[Spiegel and Zahn (1992)]{sz92} Spiegel, E.A. \& Zahn, J.P. 1992,\aap, 265,106

\bibitem[Thompson, M.J. et al. (1996)]{th96} Thompson, M.J. et al. 1996 Science, 272, 1300    
\bibitem[Tobias et al. (1998)]{to98} Tobias, S.M., Brummell, N.H., Clune, T.L. \& Toomre, J. 1998, \apj, 502, 177L
\bibitem[van Ballegoojen (1982)]{vab82} van Ballegoojen, A.A., 1982, \aap, 113, 99

\bibitem[Woosley, S.E. (1986)]{wo86} Woosley, S.E. 1986, Saas-Fee Advanced Course 16: Nucleosynthesis and Chemical Evolution I, Geneva Observatory Sauverny, Switzerland
\bibitem[Zahn (1991)]{za91} Zahn, J.P. 1991, \aap, 252, 179

  

\end{thebibliography}
\end{document}